\documentclass[icps3]{svjour}

\usepackage{epsfig}
\begin{document}
\title{Coulomb drag in the mesoscopic regime}


\titlerunning{Proceedings of {\it $19$th Nordic Semiconductor Meeting}, to appear in {\it Physica Scripta}}

\author{Niels Asger Mortensen\inst{1,2} \and Karsten Flensberg\inst{2}
  \and Antti-Pekka Jauho\inst{1}}

\authorrunning{N.A. Mortensen, K. Flensberg, and A.-P. Jauho}

\institute{Mikroelektronik Centret, Technical University of Denmark,
  \O rsteds Plads bld. 345 east, DK-2800 Kgs. Lyngby, Denmark \and \O
  rsted Laboratory, Niels Bohr Institute, University of Copenhagen,
  Universitetsparken 5, DK-2100 Copenhagen \O, Denmark}

\maketitle

\begin{abstract}
  We present a theory for Coulomb drag between two mesoscopic systems which expresses the drag in terms of scattering matrices and
  wave functions. The formalism can be applied to both ballistic and
  disordered systems and the consequences can be studied either by numerical simulations or analytic means such as perturbation theory or random matrix theory. The physics of Coulomb drag in the mesoscopic regime is very different from Coulomb drag between extended electron systems. In the mesoscopic regime we in general find fluctuations of the drag comparable to the mean value. Examples are the vanishing average drag for chaotic 2D--systems and the dominating fluctuations of drag between quasi-ballistic wires with almost ideal transmission.
\end{abstract}

\section{Introduction}\label{introduction}

Current flow in a conductor can through a Coulomb mediated drag-force
accelerate charge-carriers in a nearby conductor, thus inducing a
drag-current. The effect is active whenever the distance between the
two conductors is of the same order as the distance between the
charge-carriers -- otherwise it is suppressed by screening. In the
past years Coulomb drag in extended 2D-systems has been studied
extensively~\cite{rojo} and very recently the study of fluctuations of
the Coulomb drag has been initiated~\cite{naroa,morta,narob}. As in many other mesoscopic phenomena the fluctuations will be pronounced for temperatures
smaller than the Thouless energy. We study drag in disordered or chaotic finite-size systems with dimension much smaller than the phase-coherence length $\ell_\phi$ (see Fig.~1) and find interesting phenomena like sign-reversal, vanishing mean value, and large fluctuations of the drag response.

\begin{figure}
\begin{center}
\epsfig{file=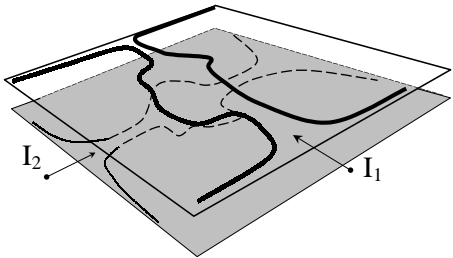, width=0.7\columnwidth,clip}
\end{center}
\caption{Schematic geometry of a mesoscopic Coulomb drag experiment.}
\label{fig:sample}
\end{figure}

\section{Formalism}\label{formalism}
Starting from the Kubo formula (as in Refs.~\cite{kamenev,flensberg}, but, here we consider systems with broken translation invariance as illustrated in Fig. 1) we calculate the drag conductance $G_{21}$ to second
order in the interaction $U_{12}$ between mesoscopic subsystems, taking the isolated systems to be otherwise non-interacting. In the dc limit~\cite{morta}
\begin{eqnarray}
G_{21}&=& \frac{e^2}{h}\int {\rm d}{\boldsymbol r}_1 {\rm d}{\boldsymbol r}_2
{\rm d}{\boldsymbol r}_1' {\rm d}{\boldsymbol r}_2'\, U_{12}({\boldsymbol
r}_1,{\boldsymbol r}_2) U_{12}({\boldsymbol r}_1',{\boldsymbol r}_2')\nonumber\\
&&\quad \times \hbar \int_{-\infty}^\infty {\rm d}\omega\,
\frac{\Delta_1(\omega,{\boldsymbol r}_1,{\boldsymbol r}_1') \Delta_2(-\omega,{\boldsymbol
r}_2,{\boldsymbol r}_2')}{2 kT\, \sinh^2(\hbar\omega/2kT)}.\label{G21}
\end{eqnarray}
Here,
\begin{eqnarray}
&&\Delta_i(\omega,{\boldsymbol r},{\boldsymbol r}')=-2i\pi^2\hbar
\sum_{\beta}\theta^i_\beta({\boldsymbol r},{\boldsymbol
r}',\varepsilon_\beta-\hbar\omega)\nonumber\\
&&\quad \times \big[n_{F}(\varepsilon_{\beta}-\hbar\omega)-
n_{F}(\varepsilon_{\beta})\big] + \big({\boldsymbol r}\leftrightarrow {\boldsymbol
r'};\, \omega\rightarrow -\omega\big), \label{Deltadef}
\end{eqnarray}
is the three point correlation function $\langle
\hat{I} \hat{\rho} \hat{\rho}\rangle$ and

\begin{equation}
\theta_\beta^i({\boldsymbol r},{\boldsymbol r}',\varepsilon)=\sum_{\alpha \gamma}
I^i_{\alpha\gamma}\rho^i_{\alpha\beta}({\boldsymbol r})
\rho^i_{\beta\gamma}({\boldsymbol
r}')\delta(\xi_\alpha)
\delta(\xi_\gamma),\label{thetadef}
\end{equation}
where $i$ labels the subsystem and $\xi_{\alpha}=\varepsilon_{\alpha} -\varepsilon$. The matrix elements are $I^i_{\alpha\gamma}=\langle \alpha|\hat{I}^i|\gamma\rangle$ and
$\rho^i_{\alpha\beta}({\boldsymbol r})=\langle \alpha|{\boldsymbol r}\rangle\langle
{\boldsymbol r}|\beta\rangle$, where $|\alpha\big>$'s are the eigenstates of
the uncoupled subsystem with energies $\varepsilon_\alpha$. With scattering states as the basis $I^i_{\alpha\beta}=\frac{\hbar}{2m}
\delta_{\varepsilon_\alpha,\varepsilon_\beta}\left(\tau^3-S^\dagger\tau^3
  S\right)_{\alpha\beta}$ is expressed in terms of the $2N\times 2N$
scattering matrix $S$~\cite{RMT} and the third Pauli matrix $\tau^3$.

Many interesting features follow immediately from eq.~(\ref{G21}) \cite{morta} but here we focus on two of them:  

\begin{itemize}

\item Phase-coherencen and lack of translation invariance leads to an arbitrary sign of the outcome of eq.~(\ref{G21}). This indicates that fluctuations might be pronounced in disordered or chaotic systems.

\item For $T\ll T_F$ the factor $\sinh^{-2}$ cuts off the frequency integration and $\Delta\propto \omega$ to lowest order. Performing the frequency integration gives $G_{21}\propto T^2$, in accordance with the usual Fermi liquid
result for electron-electron scattering.

\end{itemize}

In the following we study eq.~(\ref{G21}) in the $T^2$-regime for two cases which demonstrate fluctuations of the order of or larger than the mean value.

\begin{figure*}
\begin{center}
\epsfig{file=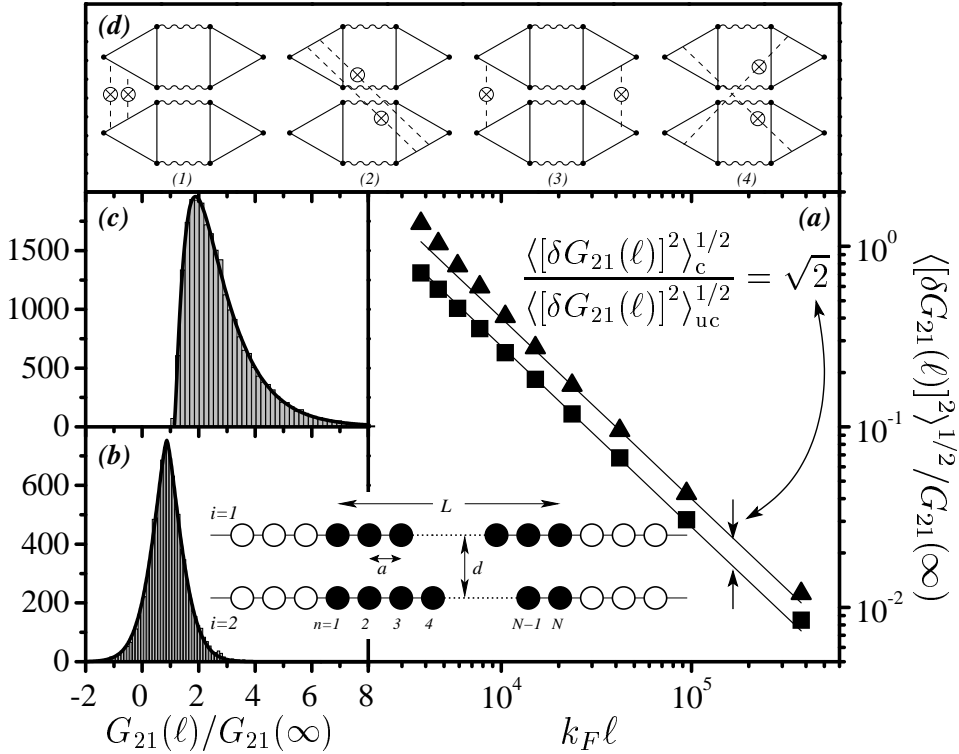, height=1.5\columnwidth,clip,angle=-90}
\end{center}
\caption{Panel (a) shows fluctuations as a function of $k_F\ell$ for $k_FL=100$ and $k_Fd=1$. The full lines are Eqs.~(\ref{1Danalytic}) and (\ref{sqrt2}) and the squares (triangles) are the numerically solution on a lattice (inset) for mutually uncorrelated (correlated) disorder. Panels (b) and (c) are the numerically obtained distributions (histograms based on $\sim 10^4$ random disorder configurations) for $k_F\ell\sim 3.8\times 10^3$ in the case of mutually uncorrelated and correlated disorder, respectively. Panel (d) shows the topologically different diagrams contributing to the fluctuations to lowest order in $1/k_F\ell$.}
\label{fig:1D}
\end{figure*}

\section{Quasi-ballistic one-dimensional wires}
We consider two weakly disordered 1D--wires of separation $d$, length $L\ll\ell_\phi$, and mean free path $\ell\gg L$. We study this system both analytically using perturbation theory and numerically by mapping the problem on to a lattice (see inset of Fig.~\ref{fig:1D}). We find that even a very small amount of disorder ($L \ll \ell$) can lead to large fluctuations for the
drag response and even reverse the sign. The origin is inter-wire interaction induced forward scattering which gives rise to a drag
response provided it is combined with disorder induced backscattering.
In contrast, in the case of clean wires the backscattering is induced
solely by the interwire interaction, and therefore the disordered case
is larger by a factor of order $\big<{\cal
  R}\big>U_{12}(0)/U_{12}(2k_F)$, with
$U_{12}(q)=\int_{0}^{L}\int_{0}^{L}dx_{1}\,dx_{2}\,e^{iq(x_{1}-x_{2})}
U_{12}(x_{1},x_{2})$ being the Fourier transformed interaction. The reflection coefficient is inversely proportional to the mean free path $\big<{\cal R}\big>\simeq L/\ell$ for $L\ll \ell$.

Assuming that the disorder potentials of the  two wires are mutually uncorrelated ($\rm uc$) \cite{kamenev,flensberg,morta} this can be shown explicitly by lowest order
perturbation theory in $1/k_F\ell$, corresponding to the
diagram $(3)$ shown in panel (d) of Fig.~\ref{fig:1D} (diagram $(1)$ gives a vanishing contribution and diagrams $(2)$ and $(4)$ are not relevant to uncorrelated case). For $k_FL\gg 1$ the result is
\begin{equation}
\frac{\big<[\delta G_{21}(\ell)]^{2}\big>_{\rm uc}^{1/2}}{G_{21}(\infty)}
\simeq\frac {\big[2\big<{\cal
R}_{1}\big>\big<{\cal R}_{2}\big>U_{12}^2(2k_{F})\widetilde{U}_{12}^{2}
(0)\big]^{1/2}}{U_{12}^2(2k_{F})},\label{1Danalytic}
\end{equation}
with $U_{12}^{2}(0)$ replaced by
\begin{eqnarray}&&\widetilde{U}_{12}^2(0)\equiv \int_{0}^{L}
    \int_{0}^{L}\int_{0}^{L}\int_{0}^{L} {\rm d}x_{1}\,{\rm d}x_{2}\,{\rm d}x_{1}'\,{\rm d}x_{2}'\, U_{12}(x_{1},x_{2})\nonumber\\
&&\quad\times U_{12}(x_{1}',x_{2}')
\Big(1-\tfrac{2|x_{1}-x_{1}'|}{L}\Big)\Big(1-\tfrac{2|x_{2}-x_{2}'|}{L}\Big).
\end{eqnarray}
The denominator is the result $G_{21}(\infty)\propto
U_{12}^{2}(2k_{F})$ for ballistic wires.  For the realistic case $U_{12}(2k_{F})\ll \widetilde{U}_{12}(0)$ it then follows that the fluctuations
of the drag can exceed the average value $\big<G_{21}(\ell)\big>_{\rm uc}\simeq G_{21}(\infty)$. This contrasts the behavior of the diagonal conductance where the fluctuations $\big<[\delta
G_{ii}]^2\big>^{1/2}$ are vanishing compared to the mean value
$\big<G_{ii}\big>= (2e^2/h)\big(1-\big<{\cal R}_i\big>\big)\sim
2e^2/h$ in the limit of weak disorder. 

The opposite limit with mutually fully correlated ($\rm c$) disorder potentials \cite{gornyi,mortb} is at first sight more complicated because the calculation of fluctuations involves also ``crossed'' diagrams like diagrams $(2)$ and $(4)$. However, for correlated disorder both
of the diagrams $(3)$ and $(4)$ contribute equally whereas for
uncorrelated disorder only diagram $(3)$ is relevant. The same is the case for diagrams $(1)$ and $(2)$ and more generally, for each topologically different diagram contributing in
the case of uncorrelated disorder there are two similar diagrams
contributing equally in case of correlated disorder (to lowest order in $1/k_F\ell$). Since symbolically $(2)=(1)$ and $(4)=(3)$ we get

\begin{equation}\label{sqrt2}
\frac{\big<[\delta G_{21}(\ell)]^{2}\big>_{\rm c}^{1/2}}{\big<[\delta G_{21}(\ell)]^{2}\big>_{\rm uc}^{1/2}}
\simeq\sqrt{\frac{(1)+(2)+(3)+(4)}{(1)+(3)}}=\sqrt{2}.
\end{equation}
Such an enhancement by correlated disorder was recently predicted for the mean drag in 2D systems \cite{gornyi}; here we find that also the fluctuations are enhanced.

Panel (a) of Fig.~\ref{fig:1D} shows a parameter-free comparison of Eqs.~(\ref{1Danalytic}) and (\ref{sqrt2}) to numerical results for $k_FL=100$ and $k_Fd=1$ in the case of bare Coulomb interaction. Both the $1/k_F\ell$-dependence, the magnitude, and the $\sqrt{2}$-enhancement are fully confirmed by the numerical simulations and similar agreement has been found for other values of $d$ and $L$ \cite{morta,mortb}. Panels (b) and (c) show typical numerically obtained distributions of $G_{21}(\ell)$ for uncorrelated and correlated disorder, respectively ($k_F\ell \sim 3.8\times 10^3$). In panel (b) the negative values correspond to sign-reversal of the drag. In panel (c) we see an enhancement of both the mean value and the fluctuations compared to panel (b). Predictions for the full distributions do not yet exist and especially the absence of sign-reversal in panel (c) is interesting. For only partially correlated disorder we numerically find enhancements of the fluctuations in the range $[1;\sqrt{2}]$ with the two limits corresponding to uncorrelated and fully correlated disorder, respectively \cite{mortb}.

\section{Chaotic mesoscopic systems}

We consider an ensemble of mesoscopic chaotic systems, such as suggested in Fig.~\ref{fig:sample}. We assume that the region where the subsystems couple by Coulomb interactions have mutually uncorrelated disorder so that (suppressing the integration variables)

\begin{eqnarray}
\big<G_{21}\big>&\propto& \int U_{12}U_{12} \big<\Delta_1\big> \big<\Delta_2\big>,\\
\big<G_{21}^2\big>&\propto& \int U_{12}U_{12} U_{12}U_{12} \big<\Delta_1\Delta_1\big> \big<\Delta_2\Delta_2\big>.
\end{eqnarray}

Starting from eq.~(\ref{Deltadef}) in the low temperature limit the task is to calculate $\big<\Delta_i\big>$ and $\big<\Delta_i\Delta_i\big>$. We do this using random matrix theory \cite{RMT} where the eigenvalues and
the wave functions are uncorrelated.

To lowest order in $1/k_F\ell$ the average of two wave functions is (see {\it e.g.} \cite{alei95})
\begin{equation}
\langle\phi_{\gamma}^{\ast}(x)\phi_{\delta}(y)\rangle\approx\delta
_{\gamma\delta}\frac{\delta\varepsilon}{\pi}\big<A(x-y)\big>, 
\end{equation}
where $\delta\varepsilon$ is the level spacing and the spectral function is given by 
\begin{equation}\label{<A>}
\langle A(r)\rangle\simeq
(m/2\hbar^2)\exp(-r/2\ell)J_0(k_{F} r).
\end{equation}
Next, consider the average
\begin{eqnarray}
&&\big< I_{\alpha\beta}\phi_{\gamma}^{\ast}(x)\phi_{\delta}
(y)\big> \propto   (\partial_{x_{1}}-\partial_{x_{2}})\\
&&\quad\quad\quad\quad\times \big<\phi_{\alpha}^{\ast}(x_{1})\phi_{\beta}(x_{2})\phi_{\gamma}^*
(x)\phi_{\delta}(y)\big> \Big|_{x_{1}=x_{2}},\nonumber
\end{eqnarray}
where to lowest order in $1/k_{F}\ell$
\begin{eqnarray}\label{4av}
&&\big< \phi_{\alpha}^*(x_{1})\phi_{\beta}(x_{2})\phi_{\gamma}^*(x)\phi_{\delta}(y)\big>   \simeq \big<\phi_{\alpha}^*(x_{1})\phi_{\beta}(x_{2})\big>\\
&&\quad\times\big<\phi_{\gamma}^*(x)\phi_{\delta
}(y)\big>  +\big<\phi_{\alpha}^*(x_{1})\phi_{\delta}(y)\big>\big<\phi_{\beta}(x_{2})\phi_{\gamma}^*(x)\big>.\nonumber
\end{eqnarray}
Due to current conservation the points $x_{1}=x_{2}$ can be anywhere. Taking them to be outside the chaotic region, the decay of the spectral function makes the second term in eq.~(\ref{4av}) vanish if $x$ or $y$ are inside. The first term amounts to performing the average over $I_{\alpha\beta}$ and $\phi_{\gamma
}^{\ast}(x)\phi_{\delta}(y)$ separately. Similar arguments hold for
higher order averages so that to lowest order in $1/k_F\ell$

\begin{eqnarray}
\frac{\big< \Delta_i(\omega,{\boldsymbol r},{\boldsymbol
r}') \big>}{\hbar\omega\,(2\pi)^2\hbar } &\simeq& {\rm Im}
\sum_{\alpha\beta\gamma}\big<
I_{\alpha\gamma}^i\big> \\
&& \times \big<\rho_{\alpha\beta}^i({\boldsymbol r})\rho_{\beta\gamma}^i({\boldsymbol r}')
\delta(\xi_\alpha^F)\delta(\xi_\beta^F)\delta(\xi_\gamma^F)\big>,\nonumber
\end{eqnarray}
and since $I$ and $\rho$ are Hermitian

\begin{eqnarray}
&&\frac{\big< \Delta_i(\omega,{\boldsymbol r},{\boldsymbol
r}') \Delta_i(\tilde\omega,{\boldsymbol s},{\boldsymbol
s}') \big>}{\hbar\omega\,\hbar\tilde\omega\,(2\pi)^4\hbar^2} \simeq \frac{1}{(2i)^2}
\sum_{\alpha\beta\gamma}\sum_{\tilde\alpha \tilde\beta\tilde\gamma}\big<\delta(\xi_\gamma^F)\delta(\xi_{\tilde\gamma}^F)\big>\nonumber \\
&&\quad \times\Big\{\big<
I_{\alpha\gamma}^iI_{\tilde\alpha\tilde\gamma}^i\big>\big< \rho_{\alpha\beta}^i({\boldsymbol r})\rho_{\beta\gamma}^i({\boldsymbol r}') \rho_{\tilde\alpha\tilde\beta}^i({\boldsymbol s})\rho_{\tilde\beta\tilde\gamma}^i({\boldsymbol s}')\delta(\xi^F) \big>\nonumber\\
&&\quad -\big<I_{\alpha\gamma}^iI_{\tilde\gamma\tilde\alpha}^i\big>\big< \rho_{\alpha\beta}^i({\boldsymbol r})\rho_{\beta\gamma}^i({\boldsymbol r}')
\rho_{\tilde\beta\tilde\alpha}^i({\boldsymbol s})\rho_{\tilde\gamma\tilde\beta}^i({\boldsymbol s}')\delta(\xi^F) \big>\nonumber\\
&&\quad -\big<I_{\gamma\alpha}^iI_{\tilde\alpha\tilde\gamma}^i\big>\big<\rho_{\beta\alpha}^i({\boldsymbol r})\rho_{\gamma\beta}^i({\boldsymbol r}')
 \rho_{\tilde\alpha\tilde\beta}^i({\boldsymbol s})\rho_{\tilde\beta\tilde\gamma}^i({\boldsymbol s}')\delta(\xi^F) \big>\nonumber\\
&&\quad +\big<I_{\gamma\alpha}^iI_{\tilde\gamma\tilde\alpha}^i\big>\big<\rho_{\beta\alpha}^i({\boldsymbol r})\rho_{\gamma\beta}^i({\boldsymbol r}')
\rho_{\tilde\beta\tilde\alpha}^i({\boldsymbol s})\rho_{\tilde\gamma\tilde\beta}^i({\boldsymbol s}')\delta(\xi^F) \big>\Big\},
\end{eqnarray}
where $\delta(\xi^F)\equiv \delta(\xi_\alpha^F)\delta(\xi_\beta^F)\delta(\xi_{\tilde\alpha}^F)\delta(\xi_{\tilde\beta}^F)$.

From the statistical properties of the $S$-matrix \cite{RMT} we
find $\big< I_{\alpha\gamma}\big> \propto
\tau^3_{\alpha\gamma}$ and since the second average in $\langle
\Delta \rangle $ is symmetric with respect to interchange of $\alpha$
and $\gamma$ we get $\langle \Delta \rangle =0 $ and thus of course $\langle G_{21} \rangle $=0. However, the fluctuations are nonzero. Since 
\begin{eqnarray}
\big< I_{\alpha\gamma} I_{\tilde\alpha\tilde\gamma}\big>&=&{\rm const.}\times \tau_{\alpha\gamma}^3\tau_{\tilde\alpha\tilde\gamma}^3\nonumber\\
&&\quad+ \left(\tfrac{\hbar}{2m}\right)^2 \big< (S^\dagger\tau^3 S)_{\alpha\gamma}(S^\dagger\tau^3
S)_{\tilde\alpha\tilde\gamma}\big>,
\end{eqnarray}
and the average in the limit of a large $N$ becomes
\\$(2N)^{-1}\delta_{\alpha\tilde\gamma}\delta_{\tilde\alpha\gamma}$ we get \cite{scatteringstates}

\begin{eqnarray}
&&\frac{\big< \Delta_i(\omega,{\boldsymbol r},{\boldsymbol
r}') \Delta_i(\tilde\omega,{\boldsymbol s},{\boldsymbol
s}') \big>}{\hbar\omega\,\hbar\tilde\omega\,(2\pi)^4\hbar^2} \simeq \frac{1}{4(2\pi)^2\hbar^2(2i)^2(2N)(2\pi)^4}
 \nonumber\\
&&\quad\quad\quad \times\Big\{\big< A_i({\boldsymbol r},{\boldsymbol s}') A_i({\boldsymbol r}',{\boldsymbol r}) A_i({\boldsymbol s},{\boldsymbol r}') A_i({\boldsymbol s}',{\boldsymbol s})  \big>\nonumber\\
&&\quad\quad\quad -\big< A_i({\boldsymbol r},{\boldsymbol s}) A_i({\boldsymbol r}',{\boldsymbol r}) A_i({\boldsymbol s},{\boldsymbol s}') A_i({\boldsymbol s}',{\boldsymbol r}')  \big>\nonumber\\
&&\quad\quad\quad -\big<A_i({\boldsymbol r},{\boldsymbol r}') A_i({\boldsymbol r}',{\boldsymbol s}') A_i({\boldsymbol s},{\boldsymbol r}) A_i({\boldsymbol s}',{\boldsymbol s}) \big>\nonumber\\
&&\quad\quad\quad +\big<A_i({\boldsymbol r},{\boldsymbol r}') A_i({\boldsymbol r}',{\boldsymbol s}) A_i({\boldsymbol s},{\boldsymbol s}') A_i({\boldsymbol s}',{\boldsymbol r}) \big>\Big\}.
\end{eqnarray}
Here, we have introduced the spectral function $A({\boldsymbol r},{\boldsymbol r}')=2\pi\sum_\alpha \phi_\alpha^*({\boldsymbol r})\phi_\alpha({\boldsymbol r}')\delta(\xi_{\alpha}^F)$. To lowest order in $1/k_F\ell$ we replace each spectral function by its average and use that $\big<A({\boldsymbol r},{\boldsymbol r}')\big>=\big<A({\boldsymbol r}',{\boldsymbol r})\big>$ so that

\begin{eqnarray}
&&\frac{\big< \Delta_i(\omega,{\boldsymbol r},{\boldsymbol
r}') \Delta_i(\tilde\omega,{\boldsymbol s},{\boldsymbol
s}') \big>}{\pi^2\,\hbar\omega\,\hbar\tilde\omega} \simeq \frac{{\cal F}_i({\boldsymbol r},{\boldsymbol
r}',{\boldsymbol s},{\boldsymbol
s}')}{2(2N)(2\pi)^4}\\
&&{\cal F}_i({\boldsymbol r},{\boldsymbol
r}',{\boldsymbol s},{\boldsymbol
s}')=\big< A_i({\boldsymbol r},{\boldsymbol r}')\big>\big< A_i({\boldsymbol s},{\boldsymbol s}')\big>,\\
&&\quad\quad\times\Big[\big< A_i({\boldsymbol r},{\boldsymbol s})\big>\big< A_i({\boldsymbol r}',{\boldsymbol s}')  \big>-\big< A_i({\boldsymbol r},{\boldsymbol s}')\big>\big< A_i({\boldsymbol r}',{\boldsymbol s}) \big>\Big].\nonumber
\end{eqnarray}
Since $\big<(\delta G_{21})^2\big>=\big<G_{21}^2\big>-\big<G_{21}\big>^2=\big<G_{21}^2\big>$ we perform the $\omega$-integration in eq.~(\ref{G21}) and get 

\begin{eqnarray}
&&\big<(\delta G_{21})^2\big>^{1/2}\simeq \frac{e^2}{h}\frac{ (kT)^2}{3\, 2^4\, N}\nonumber\\
&&\quad \times\Big[\int U_{12}({\boldsymbol
r}_1,{\boldsymbol r}_2) U_{12}({\boldsymbol r}_1',{\boldsymbol r}_2')U_{12}({\boldsymbol
s}_1,{\boldsymbol s}_2)U_{12}({\boldsymbol s}_1',{\boldsymbol s}_2')\nonumber\\
&&\quad\times{\cal F}_1({\boldsymbol r}_1,{\boldsymbol r}_1',{\boldsymbol s}_1,{\boldsymbol s}_1') {\cal F}_2({\boldsymbol
s}_2,{\boldsymbol s}_2',{\boldsymbol r}_2,{\boldsymbol r}_2')\Big]^{1/2}.
\end{eqnarray}

To obtain an estimate we note that eq.~(\ref{<A>}) is very peaked for $k_Fr<1$ and on long length scales $\big<A({\boldsymbol r})\big>\approx (\pi/4\varepsilon_F \,k_F\ell)\delta({\boldsymbol r})$ for $k_F\ell\gg 1$. Using that approximation for all spectral functions is too crude, but

\begin{eqnarray}
&&{\cal F}_i({\boldsymbol r},{\boldsymbol
r}',{\boldsymbol s},{\boldsymbol
s}')\approx\left(\frac{\pi}{4\varepsilon_F \,k_F\ell}\right)^2\big< A({\boldsymbol r},{\boldsymbol r}')\big>^2\nonumber\\
&&\quad\quad\times\big[\delta({\boldsymbol r}-{\boldsymbol s})\delta({\boldsymbol r}'-{\boldsymbol s}')-\delta({\boldsymbol r}-{\boldsymbol s}')\delta({\boldsymbol r}'-{\boldsymbol s})\big],
\end{eqnarray}
still gives a finite answer

\begin{eqnarray}
&&\big<(\delta G_{21})^2\big>^{1/2}\simeq \frac{e^2}{h}\frac{ \pi^2}{3\,2^8\, N}\frac{1}{(k_F\ell)^2}\left(\frac{kT}{\varepsilon_F}\right)^2\\
&&\quad\times\Big[\int \big< A({\boldsymbol r}_1,{\boldsymbol r}_1')\big>^2\big< A({\boldsymbol s}_2,{\boldsymbol s}_2')\big>^2\nonumber\\
&&\quad\times\big[ U_{12}({\boldsymbol
r}_1,{\boldsymbol s}_2) U_{12}({\boldsymbol r}_1',{\boldsymbol s}_2')- U_{12}({\boldsymbol
r}_1',{\boldsymbol s}_2) U_{12}({\boldsymbol r}_1,{\boldsymbol s}_2')\big]^2
\Big]^{1/2}\nonumber
\end{eqnarray}

Due to the peaked behavior of the spectral functions the mixed terms gives a vanishing contribution and introducing ${\boldsymbol u}={\boldsymbol r}_1-{\boldsymbol r}_1'$, ${\boldsymbol v}={\boldsymbol s}_2-{\boldsymbol s}_2'$, and ${\boldsymbol y}={\boldsymbol r}_1'-{\boldsymbol s}_2'$

\begin{eqnarray}
&&\big<(\delta G_{21})^2\big>^{1/2}\simeq \frac{e^2}{h}\frac{ \pi^2}{3\,2^8\,N}\frac{1}{(k_F\ell)^2}\left(\frac{kT}{\varepsilon_F}\right)^2\Big[2{\cal A}\\
&&\quad\times\int \big< A({\boldsymbol u},0)\big>^2\big< A({\boldsymbol v},0)\big>^2 U_{12}^2({\boldsymbol u}+{\boldsymbol
y},{\boldsymbol v}) U_{12}^2({\boldsymbol y},0)\Big]^{1/2},\nonumber
\end{eqnarray}
where ${\cal A}$ is the interaction area. Assuming that the screening length $r_s\ll \ell$, the range of $\big<A\big>$ is longer than the range of $U_{12}$ and we further approximate 

\begin{eqnarray}
&&\big<(\delta G_{21})^2\big>^{1/2}\simeq \frac{e^2}{h}\frac{ \pi^2}{3\,2^8\,N}\frac{1}{(k_F\ell)^2}\left(\frac{kT}{\varepsilon_F}\right)^2\sqrt{2{\cal A}}\\
&&\quad\times\Big[\int  {\rm d}{\boldsymbol v}  \big< A({\boldsymbol v},0)\big>^4\int {\rm d}{\boldsymbol u}\, U_{12}^2({\boldsymbol u},0)\int {\rm d}{\boldsymbol y}\, U_{12}^2({\boldsymbol y},0)\Big]^{1/2}.\nonumber
\end{eqnarray}
 The integral over the spectral function gives
$$
\int {\rm d}{\boldsymbol r} \big<A({\boldsymbol r},0)\big>^4=2\pi\int_0^\infty {\rm d}r\,r \big<A(r)\big>^4 =\frac{\pi}{2^7}\frac{\zeta(k_F\ell)k_F^{6}}{\varepsilon_F^4},
$$
where
$\zeta(k_F\ell)=\int_0^\infty dx\,xe^{-2x/k_F\ell}J_0^4(x)$ can be approximated by $\zeta(k_F\ell)\approx (3/2\pi^2)\ln 2k_F\ell$ for $k_F\ell\gg 1$. Similarly, for a screened interaction $U_{12}(r)=\bar{U}_{12}(r)exp(-r/r_s)$, $\bar{U}_{12}(r)=e^2/(4\pi\epsilon_0\epsilon_r r)$ we get
\begin{eqnarray}
\int {\rm d}{\boldsymbol r} U_{12}^2({\boldsymbol r},0)&=&2\pi\int_0^\infty {\rm d}r\,r {U}_{12}^2(\sqrt{r^2+d^2}) \nonumber\\
&=&2\pi \bar{U}_{12}^2(d)d^2\Gamma_0(2d/r_s),\nonumber
\end{eqnarray}
where $\Gamma_\alpha(x)=\int_x^\infty dt\, t^{\alpha-1}e^{-t}$ is the incomplete Gamma function. Collecting things we now obtain the estimate

\begin{eqnarray}
\big<(\delta G_{21})^2\big>^{1/2}&\simeq& \gamma \frac{e^2}{h}\left(\frac{kT}{\varepsilon_F}\cdot\frac{ \bar{U}_{12}(d)}{\varepsilon_F}\right)^2\nonumber\\
&&\quad\quad\quad\times\frac{\Gamma_0(\tfrac{2d}{r_s})\, d^2\,k_F\sqrt{{\cal A}}\,\ln 2k_F\ell}{\ell^2\,N},
\end{eqnarray}
where $\gamma =2^{-10}\sqrt{\pi^5/6}\simeq 0.7\times 10^{-3}$ \cite{comment2}.  Two separately contacted and electrostatically coupled quantum dots have recently been realized in a double-quantum-well GaAs/AlGaAs with $d\sim 40\, {\rm nm}$  and $\sqrt{\cal A}\sim 0.8\,{\rm \mu m}$ \cite{wilhelm}. Transport measurements indicated a strong electrostatic interaction between the dots, though, measurements of Coulomb drag where not performed. For typical numbers we estimate the drag fluctuations to be of the order of $0.1$ Ohm.

Interestingly, the above results for the mean value and fluctuations are not changed by breaking of time reversal symmetry,
in contrast to the UCF case, where the results with or without an
applied $B$-field differ by a factor of 2~\cite{RMT}.

\section{Conclusion}

We have developed a formalism for study of Coulomb drag in the mesoscopic regime which is a promising new direction for the study of mesoscopic
transport properties, since it gives an opportunity to directly study
interaction and correlation effects in mesoscopic structures. Disordered mesoscopic systems exhibit interesting and unusual physics \cite{RMT} and we have demonstrated that this is also the case in disordered Coulomb drag systems.

For quasi-ballistic 1D-wires ($L\ll\ell$) our results illustrate how the
statistics of the drag conductance depend strongly on disorder and we
find that even weak disorder can give rise to fluctuations of the same
order of magnitude as the drag conductance for the ballistic case.
This implies that the direction of drag depends on the disorder
configuration and that for a given system the sign of the drag current
will be arbitrary. Depending on its nature the presence of disorder may even enhance the average of the drag response and the fluctuations compared to absence of disorder.

For chaotic 2D systems like coupled quantum dots we find a vanishing mean drag but with finite fluctuations. These results are based on standard random matrix theory \cite{RMT} and the prediction of zero mean drag can thus be taken as a test of the degree of ergodicity of the system under investigation.


\begin{thebibliography}{}
\bibitem{scatteringstates}Since $I$ is defined for scattering states $\sum_{\gamma(\tilde\gamma)} \rightarrow (m/2\pi\hbar^2) \sum_{\gamma(\tilde\gamma)}  \int {\rm d}\varepsilon_{\gamma(\tilde\gamma)}$.
\bibitem{comment2} For $d\sim r_s$ we get $\gamma\Gamma_0(2d/r_s)\simeq 3.4\times 10^{-4}$ and with $\ln 2k_F\ell\sim 1$ this is the estimate in \cite{morta}.
\end{thebibliography}
\end{document}